# Dynamic control of spin wave spectra using spin-polarized currents

Qi Wang[1], Huaiwu Zhang[1*], Xiaoli Tang[1], Hans Fangohr[2], Feiming Bai[1], Zhiyong Zhong[1**]

[1]*State key Laboratory of Electronic Thin Films and Integrated Devices, University of Electronic Science and Technology of China, Chengdu, 610054, China*

[2]*Faculty of Engineering and the Environment, University of Southampton, Southampton SO17 1BJ, United Kingdom*

We describe a method of controlling the spin wave spectra dynamically in a uniform nanostripe waveguide through spin-polarized currents. A stable periodic magnetization structure is observed when the current flows vertically through the center of nanostripe waveguide. After being excited, the spin wave is transmitted at the sides of the waveguide. Numerical simulations of spin-wave transmission and dispersion curves reveal a single, pronounced band gap. Moreover, the periodic magnetization structure can be turned on and off by the spin-polarized current. The switching process from full rejection to full transmission takes place within less than 3ns. Thus, this type magnonic waveguide can be utilized for low-dissipation spin wave based filters.

** zzy@uestc.edu.cn
* hwzhang@uestc.edu.cn

A Magnonic Waveguide (MW) consists of periodic magnetic structures. The periodic structure affects the spin wave dispersion curve by creating forbidden bands at the Brillouin zone boundaries due to Bragg reflection. In recent years, the characteristics of spin wave propagation in MWs, which were fabricated by different magnetic materials [1,2] and different widths [3,4], were investigated in detail. However, the spectra of spin waves in these MWs cannot be changed dynamically after its fabrication. Dynamic artificial magnonic crystals are currently the focus of much interest because the spin wave spectra in these magnonic cyrstals can be modulated dynamically. It was reported that the spin wave spectra could be dynamically modulated by local Oersted fields produced by electrical currents on the surface of low-loss Yttrium iron garnet (YIG) [5-6]. Nevertheless, MWs based on metal magnetic thin films are more suitable to integrate with CMOS circuits. Recently, Volkov *et al.* have studied the action of the strong perpendicular spin-polarized current on ferromagnetic systems for two-dimensional films [7-9] and a narrow one-dimensional wire [10]. In the both case, the stable periodic magnetization structures induced by spin-polarized current were found in the both case.

In this letter, we present a way to modulate spin wave spectra in a uniform waveguide by spin-polarized current. The waveguide considered in this letter is presented in Fig. 1. The length, width and thickness of the stripe are L = 2000 nm, w = 160 nm and h = 10 nm, respectively. The width of the pinned layer is $\omega_p$ = 30 nm. A thin nonmagnetic spacer is placed between the pinned layer and the stripe. The damping constant is $\alpha$ = 0.01. In order to suppress spin wave reflections, the damping parameter is increased to 0.5 at the ends of the waveguide (x < 20 nm and x > 1980 nm). The spin-polarized current flows along the perpendicular direction to the center of the nanostripe waveguide. In order to excite spin wave within a wide frequency range, a sinc field pulse, ranging from 0 to 60 GHz, was applied to the y-axis. A static bias magnetic field $H_0$ = 300 Oe was applied along the x-axis to avoid forming domain wall [11]. Our micromagnetic simulation is based on the Landau-Lifshitz-Gilbert equation with the Slonczewski-Berger spin torque term [12-15].

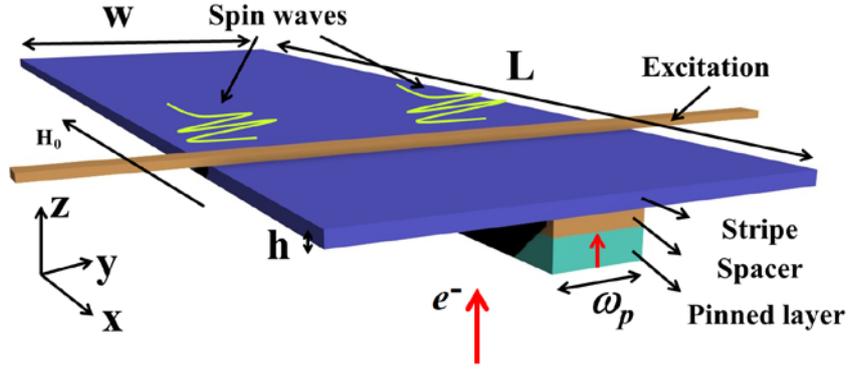

Fig. 1(Color online) Schematic view of the three-layer waveguide. The spin polarized current flows along z-direction, i.e. perpendicular to the stripe. An alternating excitation field applied along the y-axis generates spin wave along the x-axis and a static bias magnetic field $H_0 = 300$ Oe is applied along the x-axis. The length, width and thickness of stripe are L=1200 nm, w=160 nm and h=10 nm, respectively. The width of pinned layer is $\omega_p = 30$ nm. A thin nonmagnetic spacer is placed between the pinned layer and stripe.

Figure 2 shows the spatial distributions of magnetization of MW under different conditions. In the absence of a spin-polarized current, the relaxed spatial distribution of magnetization is uniform in the waveguide as shown in Fig. 2(a). However, when the spin-polarized current is applied on the nanostripe waveguide, the spin transfer torque produced by the spin-polarized current creates a magnetization structure in the center of the waveguide that is periodic in the x-direction [see Fig. 2(b)]. This periodic structure consists of two chains of vortex-antivortex configurations, and is similar to the quasicrystal state shown in Fig 2(f) in Ref. 9. We show the z-component of the magnetization, $M_z$, in the system along a particular line scanned along the x-axis and y-axis in Fig. 2 (c) and (d), respectively. While the magnetization is uniform along both the long and short axis without the spin-polarized current (Fig 2(a)), the magnetization distribution is non-uniform with the spin-polarized current applied (Fig 2(b-d)): the spin transfer torque creates a periodic magnetization structure in the middle of the nanostripe waveguide. The magnetization is almost parallel to the z-axis in the middle of nanostripe waveguide in a region centered around the location where the spin-torque term is applied. The spin waves are not allowed to propagate in this region, because the spin-polarized current contributes an effective damping, which is usually greater than the natural one [7,9]. Hence, we focus on the edge of the nanostripe waveguide where spin waves are allowed to propagate. The effective width of this channel is about 60 nm, see Fig. 2(d). Nevertheless, the z component of magnetization along the length direction is oscillating in this channel caused by exchange and dipolar interaction. These oscillations of the magnetization component $M_z$ as a function of length (x-axis) are plotted in Fig. 2(c). From the figure, one can obtain the periodicity D = 60 nm when spin-polarized current was applied.

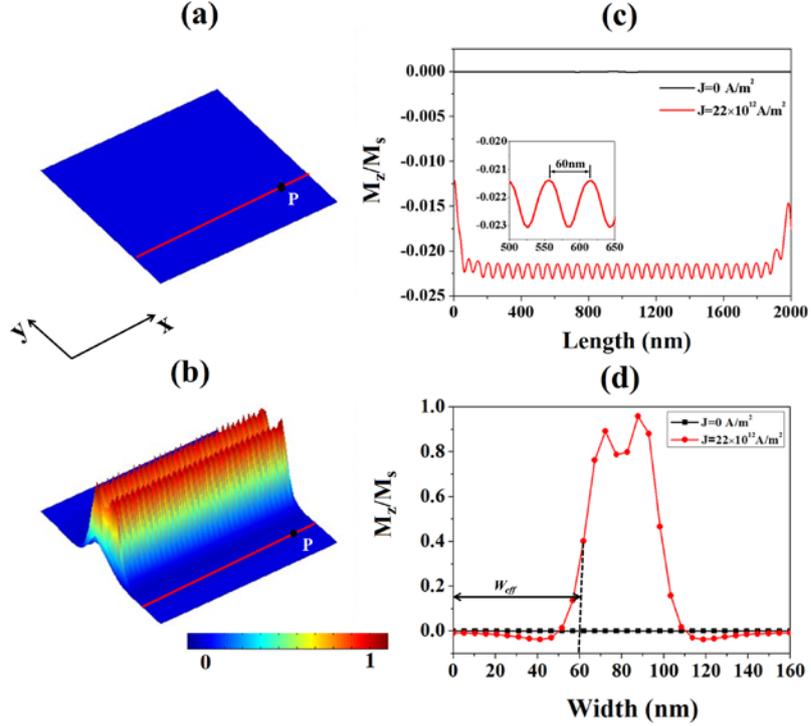

Fig. 2 (Color online) (a), (b) The remnant magnetization ($M_z/M_s$) of stripe for (a) J = 0 (b) J = 22×10$^{12}$ A/m$^2$, respectively. The z component of magnetization $M_z$ is extracted along (c) x-axis (red line in Fig2. (a), (b)) and (d) y-axis, respectively.

The spin wave spectra, characters of transmission and dispersion curves are shown in Fig. 3. The spectra are obtained through the fast Fourier transform (FFT) of the temporal $M_z/M_s$ along the red line shown in Fig. 2(a). The frequency spectra clearly reveal allowed and forbidden bands: low values represented by blue/green represent forbidden bands and high values associated with red and orange are the allowed bands. Fig. 3(a-c) show that for zero spin-polarized current, spin wave transmission is suppressed below 7.5 GHz [16], which originates from the nanostripe width confinement [17,18]. There is no band gap above 7.5 GHz in the spectra. Fig. 3(d-f) show that the spin-polarized current changes the spin-wave spectra drastically: A new gap emerges above 7.5 GHz as can be seen in Fig. 3(d). The transmission characteristic obtained by integrating the spin-wave intensity from 1400 nm to 1600 nm is displayed in Fig. 2(e), also showing the new spin-wave band gap that appears in the presence of the spin-polarized current. By calculating the dispersion curves numerically and analytically [17-19] for spin-wave [analytical result shown as black line in Fig. 3(c), and (f)], we find that the center frequency of the band gap is closely correlated with the first order Bragg reflection wavenumber: the center frequency of 13.2 GHz is identified to belong to the first width mode and is formed at the Bragg wavenumber $k_{D1} = \pi/D = 0.052$ nm$^{-1}$, where D is the periodicity of magnetization. Other higher frequency band gaps are not found in the spectrum, because the spin-polarized current induced magnonic crystal is designed in such a way that variations of the magnetization along the length direction are practically sinusoidal as shown in Fig. 2(c). Therefore the spectrum of this MW contains only one band gap and this phenomenon has been observed in experiment and theory [5,20,21].

Fig. 3 demonstrates the key result of this letter: a spin-polarized current can induce band gaps in a uniform nanostripe waveguide, opening the path to electrically switchable band gaps. Further studies show that the center frequency and depth of the band gap is almost stable with increasing the spin-polarized current density from $22\times10^{12}$ to $24\times10^{12}$ A/m$^2$ (data not shown).

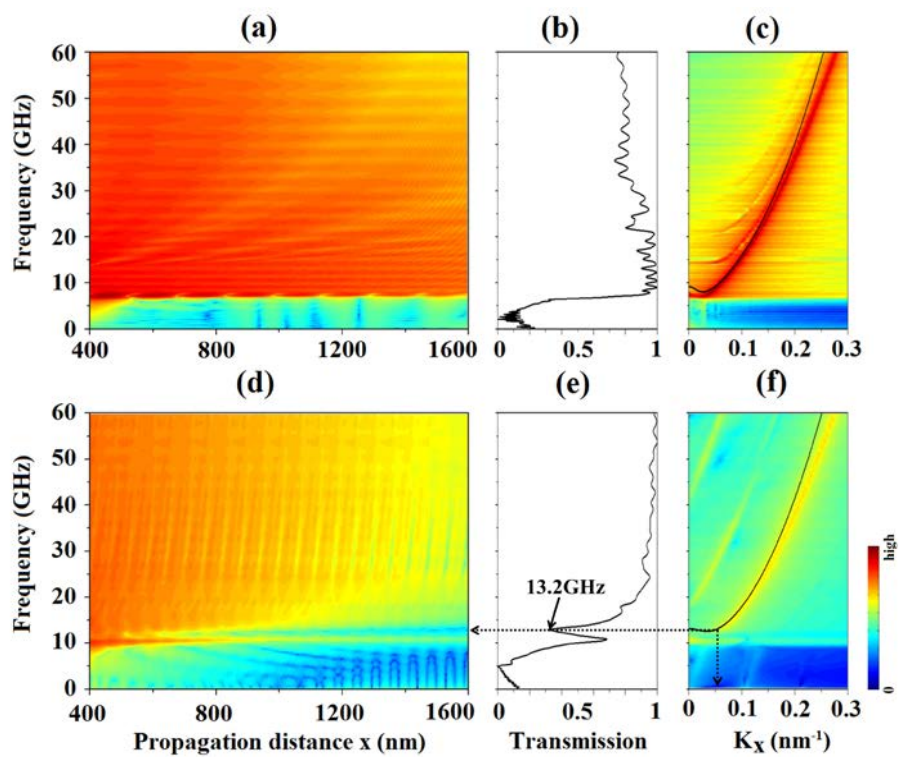

Fig. 3 (Color online) Character of MW without spin-polarized current (top row, plots (a) to (c)) and with spin-polarized current J=$22\times10^{12}$A/m$^2$ (bottom row, plots (d) to (f)). Figures 2(a) and (d) show frequency spectra obtained from FFTs of $M_z/M_s$ along the red line shown in Fig.2(a). Figures 2 (b) and (e) show spin-wave transmission characteristics obtained by integrating the spin-wave intensity of the MW (from 1400nm to 1600nm). The inset frequencies marked the center frequency of the band gap. Figures 2 (c) and (f) show the dispersion relation of the first modes obtained by micromagnetic simulation and analytical

calculation (the black line). The black dotted lines denote the positions of the switchable band gap and the corresponding wave vectors $k_x$.

The results above show that it is possible to switch the periodic structure on and off and realize dynamic control of spin wave spectra through spin-polarized currents. We now focus on the switching time required for this process. A signal with a frequency $f = f_{gap} = 13.2$ GHz was applied. The continuous spin wave signal was picked up at P point as shown in Fig. 2(a), (b). A spin-polarized current ($22 \times 10^{12}$ A/m$^2$) was applied during the first 5 ns and then it was set to zero for the next 10 ns as shown in the inset of the Fig. 4(a). The z-component of the magnetization at point P of the transmitted spin wave signal is shown in Fig. 4(a) (black line). There is no spin wave signal to be detected during the first 5 ns: the transmission of the spin wave is effectively suppressed by the spin-polarized current. After about 8 ns, by contrast, we observe the periodic spin wave signal. The Fourier spectrum, based on data from 8 to 15ns, shows that the main frequency of the spin wave is equal to the frequency of the excitation signal. We also observe the second and third harmonic, which is one of the most the universal phenomena appearing in nonlinear systems [22-25]. The z-component of magnetization shows excitation and subsequent decay between 5 ns and 8 ns. In order to elucidate the physical origin of this, we also measure the magnetization component $M_z$ at P point without the excitation signal as shown in Fig.4 (red line), which reveals a similar patterned. Therefore, we conclude that the attenuation is due to spin-polarized current switching and subsequent relaxation of the magnetization pattern rather than due to spin wave. It takes about 4ns for the spin-wave propagation to be fully established after the current is switched off. Fig. 4(b) shows results of the corresponding process of switching the spin-polarized current on after 5ns: the transmission of the spin wave is effectively suppressed by the spin-polarized current, and the transition of full transmission to full rejection of spin-waves also takes approximately 3 ns.

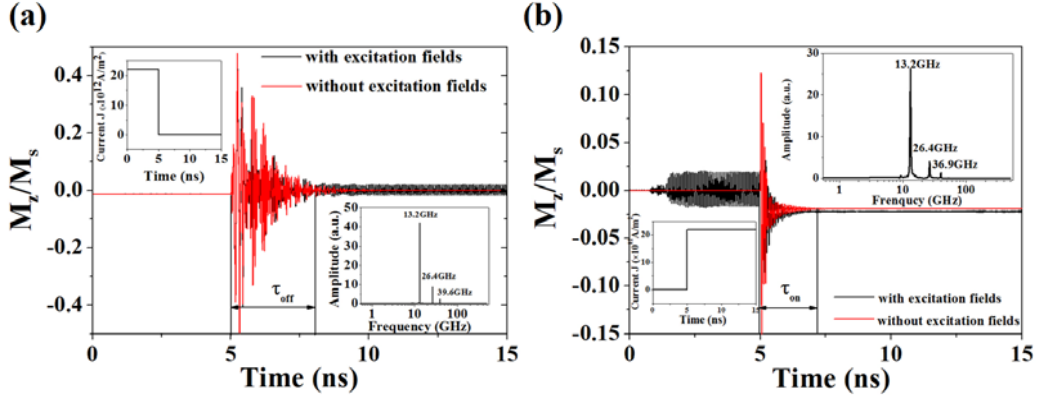

Fig. 4 (Color online) Time-dependent spin wave amplitude at P point with excitation signal (black line) and without excitation signal (red line). The insets show the spin-polarized current as a function of time and the Fourier spectrum of out-of-plane magnetization component $M_z$. Plot (a) shows the spin-polarized current being switched off after 5ns, and (b) shows the opposite process of switching the current on after 5ns. The results demonstrate that the switchable band gap can be activated and de-activated within 3ns.

In order to study the effect of the spin-polarized current on the spin waves with frequencies far away from the center frequency of the band gap $f_{gap}$ = 13.2 GHz, we applied a signal frequency $f$ = 20GHz. As seen from Fig. 5, the spin wave is not suppressed by the spin-polarized current and completely passes through the waveguide. The spin-polarized current can effectively suppress the spin waves whose frequencies are nearby the center frequency of the band gap. However, the spin-polarized current has little impact on the spin wave with other frequencies.

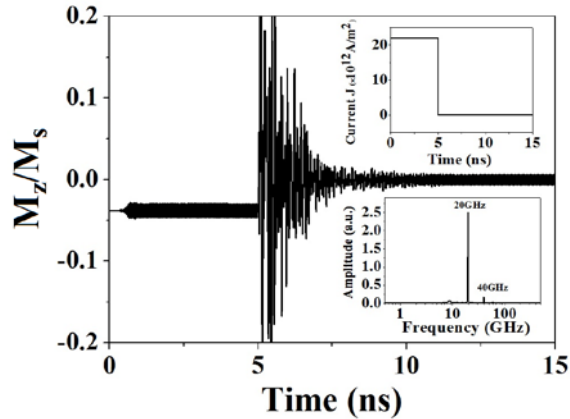

Fig. 5 Time-dependent spin wave amplitude at P point with excitation signal ($f$ = 20 GHz). The insets show the spin-polarized current as a function of time and the Fourier spectrum of out-of-plane magnetization component $M_z$. The data shows that spin-waves with frequencies away from the band-gap are not affected by the changes of the magnetization texture originating from the spin-polarised current.

In this letter, we have presented a dynamic magnonic crystal design in a uniform waveguide that can be switched electrically using a spin-polarized current. The spin-polarized current induces periodic magnetization structure in the uniform waveguide which leads to a pronounced spin wave band gap. This periodic magnetizaton structure is dynamically controllable, i.e. it is possible to switch the spin waves transmission at band gap frequencies on and off, using well established technology of electric currents. This type of magnonic crystal can be used for low-power filters and other magnonic devices.

**Acknowledgements**

This paper is supported by the National Nature Science Foundation of China under Grant Nos. 61271037, 51171038, 51132003, and 61271038, the SRFDP under No 20120185110029, and Key Technology R&D Program of Sichuan No. 2013GZ0025.

# Reference

[1] S. Neusser and D. Grundler, Adv. Mater. 21, 2927(2009)

[2] Z. K. Wang, V. L. Zhang, H. S. Lim, S. C. Ng, M. H. Kuok, S. Jain, and A. O. Adeyeye, Appl. Phys. Lett. 94, 083112(2009)

[3] K. –S. Lee, D. –S. Han, and S. –K. Kim, Phys. Rev. Lett. 102, 127202(2009)

[4] S. –K. Kim, K. –S. Lee, and D. –S. Han, Appl. Phys. Lett. 95, 082507(2009)

[5] A. V. Chumak, T. Neumann, A. A. Serga, B. Hillebrands, and M. P. Kostylev, J. Phys. D: Appl. Phys. 42, 205005(2009)

[6] A. D. Karenowska, V. S. Tiberkevich, A. N. Slavin, A. V. Chumak, A. A. Serga, and B. Hillebrands, Phy. Rev. Lett. 108, 015505(2012)

[7] Oleksii M. Volkov, Volodymyr P. Kravchuk, Denis D. Sheka, and Yuri Gaididei, Phys. Rev. B 84, 052404(2011)

[8] Yuri Gaididei, Oleksii M. Volkov, Volodymyr P. Kravchuk, and Denis D. Sheka, Phys. Rev. B 86, 144401(2012)

[9] Oleksii M. Volkov, Volodymyr P. Kravchuk, Denis D. Sheka, Franz G. Mertens, and Yuri

Gaididei, Appl. Phys. Lett. 103, 222401(2013)

[10] Volodymyr P. Kravchuk, Oleksii M. Volkov, Denis D. Sheka, and Yuri gaididei, Phys. Rev. B 87, 224402(2013)

[11] In general, the magnetization favors alignment parallel to the long axis of the structure due to the shape anisotropy. However, in our study the spin-polarized current excites significant magnetization dynamics. In particular when switching the spin-polarized current off, the magnetization may relax into undesired non-uniform configurations with domain separated by a domain wall.

[12] J. C. Slonczewski, J. Magn. Magn. Mater. 159, L1(1996)

[13] J. C. Slonczewski, J. Magn. Magn. Mater. 247, 324(2002)

[14] J. Xiao, A. Zangwill, and M. D. Stiles, Phys. Rev. B 70, 172405(2004)

[15] We use the OOMMF code, version 1.2a5 for material parameters of Permalloy: saturation magnetization $M_s = 8.6\times10^5$ A/m, exchange constant $A = 1.3\times10^{11}$ J/m. the cell size is $5\times5\times10$ nm. The current parameters are the following: polarization degree P=0.4 and resistance mismatch between the spacer and the ferromagnet stripe $\Lambda$=2.

[16] The widths of the intrinsic potential barriers are different because of the width of the spin wave channels change when the current apply.

[17] T. W. O`Keeffe and R. W. Patterson J. Appl. Phys. 49, 4886(1978)

[18] S. Choi, K. –S. Lee, K. Yu. Guslienko, and S. –K, Kim, Phys. Rev. Lett. 98, 087205(2007)

[19] The parameters used for analytical calculations were as follows: the exchange constant $A = 13\times10^{-12}$, the saturation magnetization $M_s = 8.6\times10^5$ A/m, the effective width of channel $w_{eff} = 60$ nm, and zero internal field.

[20] F. Ciubotaru, A. V. Chumak, N. Yu. Grigoryeva, A. A. Serga and B. Hillebrands, J. Phys. D: Appl. Phys. 45, 255002(2012)

[21] A. V. Chumak, V. S. Tiberkevich, A. D. karenowska, A. A. Serga, J. F. Gregg, A. N. Slavin, and B. Hillebrands, Nat. Commun. 1, 141(2010)

[22] V. E. Demidov, M. P. Kostylev, K. Rott, P. Krysteczko, G. Reiss, and S. O. Demokritov, Phys. Rev. B, 83, 054408(2011)

[23] T. Sebastian, T. Brächer, P. Pirro, A. A. Serga, B. Hillebrands, T. Kubota, H. Nagauma, M. Oogane, and Y. Ando, Phys. Rev. Lett., 110, 067201(2013)

[24] J. Marsh, V. Zagorodnii, Z. Celinski, and R. E. Camley, Appl. Phys. Lett., 100, 102404(2012)

[25] V. E. Demidov, H. Ulrichs, S. Urazhdin, S. O. demokritov, V. Bessonov, R. Gieniusz, and A. Maziewski, Appl. Phys. Lett., 99, 012505(2011)